# High $E_T$ Jet physics at the Tevatron

*Thomas Nunnemann*[1,2]

(1) *Ludwig-Maximillians Universität, Munich, Germany, e-mail:* <u>Thomas.Nunnemann@lmu.de</u>
(2) *for the D0 and CDF collaborations*

**Abstract**

Recent measurements of high $E_T$ jet production at the Tevatron are presented. These data provide stringent tests of perturbative QCD as well as constraints on parton distribution functions, the strong coupling constant and models implemented in event generators. Measurements of inclusive jet and multijet production as well as the associated production of jets with the electroweak bosons obtained by both the D0 and CDF collaborations are reviewed.

**Introduction**

Measurements of jet production in proton-antiproton collisions at the Tevatron provide stringent tests of quantum chromodynamics (QCD). At large transverse momentum $p_T$ the production of jets can be calculated in perturbative QCD (pQCD). Therefore measurements of jet production are directly sensitive to both the strong coupling constant $\alpha_s$ and the parton distribution functions (PDFs) of the proton. In particular, the gluon PDF is constrained by inclusive jet cross section measurements, while the associated production of electroweak bosons with heavy quark jets provides constraints on $b$ and $c$ quark PDFs.

Jet production, in particular when associated with a $W$ and $Z$ boson, is an important background for searches for the Higgs boson, supersymmetry, or other new phenomena. As processes with large multiplicities can only be calculated at leading order (LO) in pQCD with unavoidable large uncertainties, their precise measurement is essential to tune event generators used at the Tevatron and the LHC.

Some extensions of the standard model, e.g. models predicting excited quarks or quark substructures, would provide distinct signatures with jets, e.g. a resonance in the dijet mass spectrum.

This report is organized as follows: First, measurements of inclusive jet production and their constraints on the gluon PDF and $\alpha_s$ are presented, followed by a summary of measurements of multijet production and jet shapes. The last sections review measurements of the associated production of jets with the electroweak bosons with an emphasis on the production of heavy flavour jets.

**Inclusive jet production**

At the Tevatron, measurements of jet production are mostly based on jets reconstructed using the Run II midpoint cone algorithm. To allow a comparison with theory, jet measurements are corrected to particle (hadron) level, while predictions from pQCD at the parton level need to be corrected for fragmentation, hadronization and the underlying event.

The correction of the data relies on a precise determination of the jet energy scale calibration, which is in most cases the source of the dominant systematic uncertainty. The jet energy scale calibration is mostly based on a determination of the calorimeter response using the $E_T$ balance in $\gamma$+jet events.

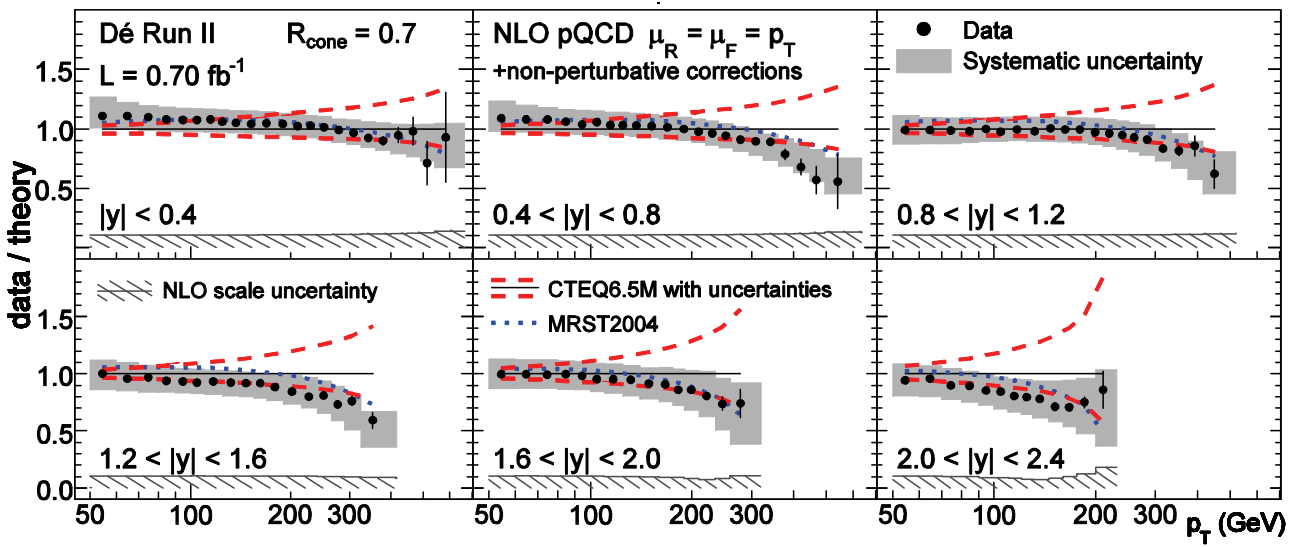

**Fig. 1:** *The D0 measurement of the inclusive jet cross section divided by the NLO QCD prediction as a function of jet $p_T$ in six $|y|$ bins [2].*

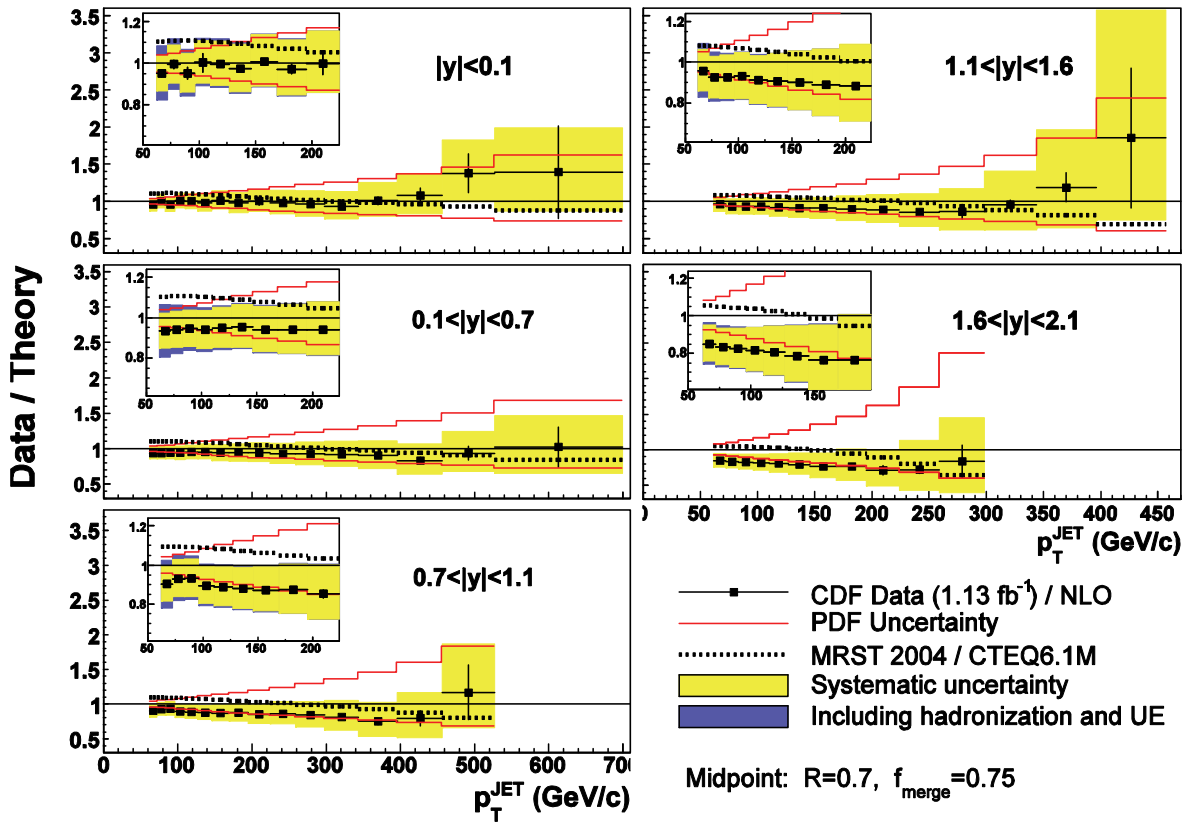

**Fig. 2:** *The CDF measurement of the inclusive jet cross section divided by the NLO QCD prediction as a function of jet $p_T$ in five $|y|$ bins [1].*

Additional corrections account for offset energy (from e.g. noise or pileup of additional collisions) and the net flow of energy in and out of the cone due to showering. The D0 collaboration reached uncertainties on the jet energy scale as low as 1.2% for central jets with $p_T$ between 100-300 GeV.

Both, the CDF and D0 collaborations measured the double differential cross section $\frac{d^2\sigma}{dp_T \, dy}$ for inclusive jet production in proton-antiproton collisions at $\sqrt{s} = 1.96$ TeV as function of transverse momentum $p_T$ and rapidity $y$ using data sets with integrated luminosities of 1.13 fb$^{-1}$ and 0.7 fb$^{-1}$, respectively [1,2]. While these measurements use the midpoint cone algorithm with a cone size of $R = 0.7$, the CDF collaboration in addition measured the inclusive jet cross section based on the $k_T$ algorithm [3]. These measurements reach jet rapidities of up to $y = 2$ and transverse momenta of up to $p_T = 600$ GeV. Figures 1 and 2 show comparisons of the D0 and CDF data, respectively, with the next-to-leading order (NLO) pQCD predictions [4]. The measurements agree with theory, although the predictions based on the CTEQ6.5M and CTEQ6.1M PDF sets, respectively, are slightly higher than the data at large jet $p_T$, indicative of a too hard gluon distribution at large $x$ in these parameterizations as $x \propto \frac{2p_T}{\sqrt{s}}$. The MRST2004 PDF set is in better agreement with both measurements.

Over a wide kinematic range, the experimental uncertainties, which are dominantly due to the jet energy scale, are smaller than the uncertainties on the theory prediction which mostly reflect the uncertainties of the PDFs. Therefore these datasets should provide improved constraints on the gluon PDF, in particular at large $x$. In fact, the recent global PDF fits MSTW2008 [5] and CT10 [6] which include the recent inclusive jet measurements predict a softer gluon PDF at large $x$ compared to their previous fits.

The first measurements of inclusive jet cross sections in proton-proton collisions at $\sqrt{s} = 7$ TeV at the LHC have recently been presented by the ATLAS and CMS collaborations [7, 8]. These measurements still have a limited kinematic reach at large jet $p_T$ and sizeable uncertainties, since the initial data sets include only integrated luminosities of up to 60 nb$^{-1}$ and since the first calibrations of the jet energy scale still have large uncertainties. With increasing luminosities and more refined calibrations, a substantial improvement in precision and kinematic reach is expected soon. Nevertheless, the Tevatron experiments will still be more sensitive to the gluon PDF at high $x$ for several years as the production cross section for central jets at large $x_T = \frac{2p_T}{\sqrt{s}} \propto x$ is substantially larger at the Tevatron compared to the LHC.

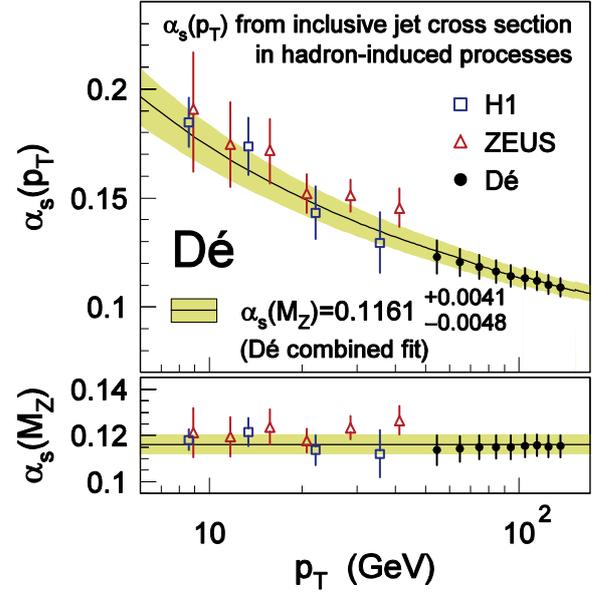

*Fig. 3:* *The D0 measurement of $\alpha_s(p_T)$ (top) and $\alpha_s(M_Z)$ (bottom) [9]. For comparison, HERA DIS jet data and the result of the fit to the D0 data are shown in addition.*

Based on their measurement of the inclusive jet production cross section, the D0 collaboration determined the running of the strong coupling constant $\alpha_s$ with the scale given by the jet $p_T$ [9]. The inclusive jet cross section is directly related to $\alpha_s$ and its pQCD prediction can be written as a convolution of the sum of the perturbative coefficient functions $c_n$ and the PDFs $f_{1,2}$ of the incoming hadrons: $\sigma_{pert}(\alpha_s) = (\sum_n \alpha_s^n c_n) \otimes f_1(\alpha_s) \otimes f_2(\alpha_s)$.

While conceptually PDFs are only functions of the momentum fraction $x$ and the scale $\mu_F$, in practice their parameterizations implicitly depend on $\alpha_s$ via the observables included in the PDF fits.

The D0 analysis uses $c_n$ given at NLO with two-loop corrections and the $\alpha_s$ dependent PDF parameterizations of the MSTW2008 NNLO fit.

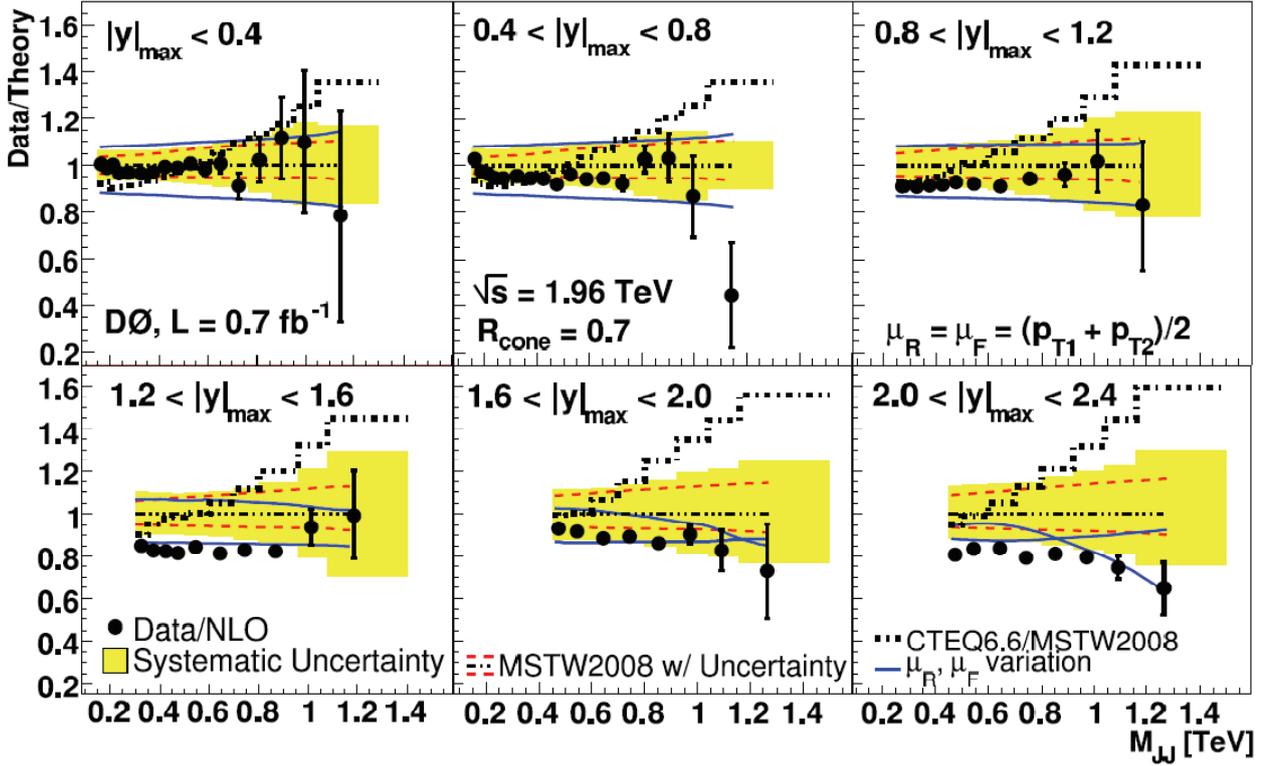

**Fig. 4:** *Measured dijet production cross section as function of the dijet invariant mass $M_{jj}$ divided by the NLO QCD prediction [10].*

To minimize the correlation of the PDF fit with the jet cross section measurement, the analysis uses only the data points in a kinematic region corresponding to low $x$. The running of $\alpha_s$ at the highest $p_T$ (see Fig. 3) was extracted together with the most precise determination of the strong coupling constant at a hadron collider: $\alpha_s(M_Z) = 0.1161^{+0.0041}_{-0.0048}$.

**Dijet production**

A measurement of the dijet production cross section as function of the dijet invariant mass $M_{jj}$ can not only be used to test pQCD and constrain PDFs, but also to search for dijet mass resonances as signatures of physics beyond the standard model (SM).

The D0 collaboration measured the inclusive dijet double differential cross section as a function of $M_{jj}$ and of the largest absolute rapidity ($y_{\max}$) of the two jets with the largest $p_T$ in an event using 0.7 fb$^{-1}$ of data [10]. Dijet events with $y_{\max} < 2.4$ and $M_{jj}$ up to 1.3 TeV were recorded. Whereas for central rapidities a good agreement between the data and the NLO pQCD prediction based on the MSTW2008 PDF set was found (Fig. 4), the observed rate of dijet events with a forward jet was lower than predicted. The discrepancies were found to be considerably larger when the CTEQ6.6 PDFs were used in the calculation.

Several extensions of the standard model predict the existence of new particles which can decay into dijets. These high mass resonances can be produced by *s*-channel annihilation and thus lead to dijet signatures with mostly central jets.

The CDF collaboration searched for narrow mass resonances in the dijet mass distribution requiring jet rapidities $|y| < 1$ [11]. The observed dijet mass distribution and the expected signal for an hypothetical excited quark $q^*$ with an assumed mass ranging from 300-1100 GeV are shown in Fig. 5. No indication of a resonant signal was found and mass limits for various new hypothetical particles were set; e.g. excited quarks with masses below 870 GeV were excluded.[1]

The ATLAS collaboration recently published a similar search based on an integrated luminosity of 315 nb$^{-1}$ which improves the $q^*$ mass limit to 1.26 TeV despite the small size of the data set [12], which impressively demonstrates the largely increased sensitivity at the LHC due to the higher collision energy.

---

[1] In this report, all limits are given at 95% C.L.

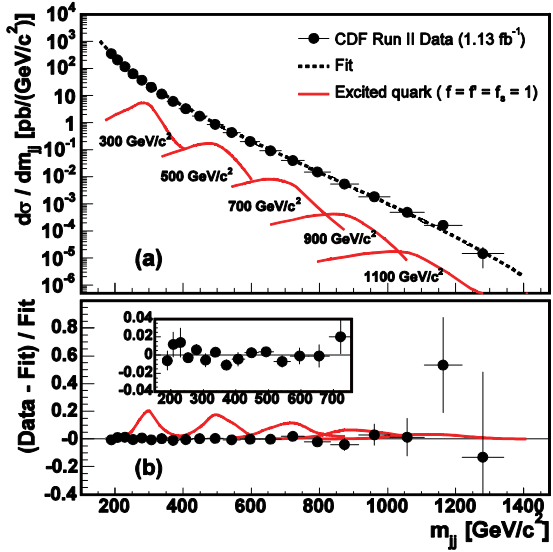

**Fig. 5:** *(a) The CDF measurement of the dijet mass spectrum shown with the predicted signals from excited quarks $q^*$. (b) The fractional difference between the measurement and a fit to the data compared to the $q^*$ signal [11].*

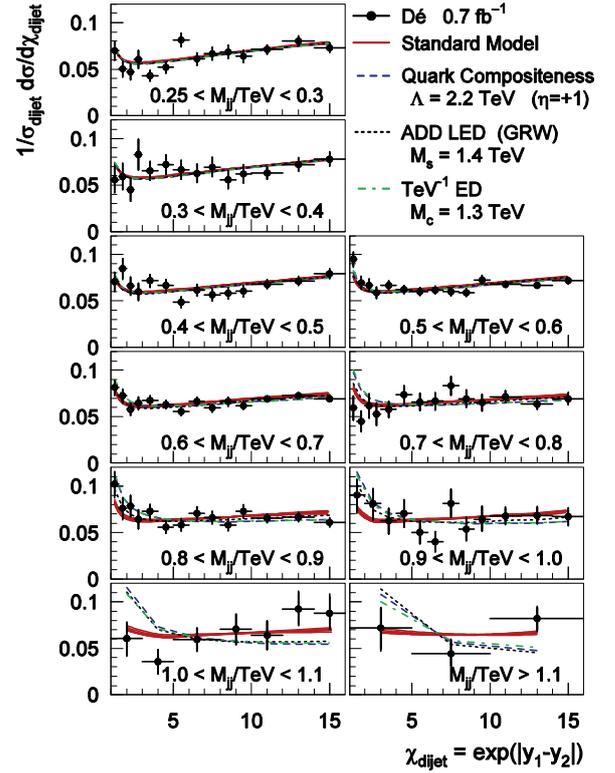

**Fig. 6:** *D0 measurement of the $\chi_{\text{dijet}}$ distribution in different bins of dijet mass $M_{jj}$ compared to the SM prediction and to the expectation of SM extensions with quark compositeness or extra dimensions [13].*

New physics beyond the SM can also modify the dijet angular distribution. A particular sensitive observable is $\chi_{\text{dijet}} = \exp(|y_1 - y_2|)$, defined such that the differential cross section $d\sigma/\chi_{\text{dijet}}$ is independent of $\chi_{\text{dijet}}$ for $t$-channel Rutherford scattering. Whereas for QCD dijet production the dependence on $\chi_{\text{dijet}}$ is nearly flat, new physics processes would result in an excess at small $\chi_{\text{dijet}}$ (corresponding to large scattering angles in the centre-of-mass-system) and large $M_{jj}$.

The D0 collaboration measured the normalized differential cross section $1/\sigma_{\text{dijet}}\, d\sigma/\chi_{\text{dijet}}$ in bins of $M_{jj}$ ranging from 0.25 to more than 1.1 TeV using their 0.7 fb$^{-1}$ data set [13]. The data were found to be consistent with the predictions of NLO pQCD (Fig. 6) and limits on new physics contributions were derived: Models predicting quark compositeness with energy scales below about 3.0 TeV were excluded and for various models with extra spatial dimensions limits on the effective Planck or compactification scale of about 1.6 TeV were set.

Recently, the ATLAS collaboration published a limit of 3.4 TeV on the quark compositeness scale based on the dijet $\chi_{\text{dijet}}$ distribution measured in a dataset with an integrated luminosity of 3.1 pb$^{-1}$ [14], which supersedes the previously most stringent limit from the D0 experiment.

## Trijet production

Three jet production provides a further test of QCD which is less dependent on the parameterization of PDFs and in principle directly probes the strong coupling constant $\alpha_s$.

Using their well established 0.7 fb$^{-1}$ data set, the D0 collaboration presented a preliminary measurement of the three-jet mass distribution $d\sigma/dM_{\text{3jet}}$ reaching masses up to 1.1 TeV [15]. The data were found to be reasonably described by NLO pQCD predictions based on MSTW2008 PDFs, but similar to dijet production, the data prefer the lower bound on the theory prediction.

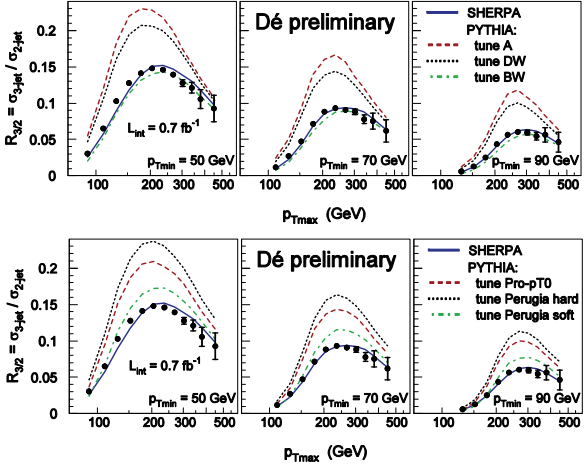

**Fig. 7:** *The D0 measurement of the ratio $R_{3/2}$ of trijet and dijet cross sections, measured as function of the leading jet $p_T$ ($p_{T\max}$) for different $p_{T\min}$ requirements for the other jets [16]. The data are compared to the predictions of Sherpa and Pythia using different tunes based on the $Q^2$ ordered (top) and $p_T$ ordered (bottom) parton shower.*

The D0 collaboration also presented a preliminary measurement of $R_{3/2}$, the ratio of inclusive trijet and dijet production cross sections, as function of the transverse momentum $p_{T\max}$ of the leading jet [16]. The measured ratio when requiring a minimal transverse momentum for the other jets ($p_{T\min}$) is shown in Fig. 7. Using default settings, the Sherpa event generator [17], which incorporates a matching of matrix elements and parton showers based on the CKKW prescription, provides a good description of the data. Predictions of the Pythia event generator [18] using different tunes (based on both the $Q^2$ and $p_T$ ordered parton shower models) vary considerably.

All Pythia tunes except BW predict a too high rate of trijet events. The tune BW is known not to describe the dijet azimuthal decorrelation [19]. Given the large dependence of the predictions on non-perturbative model parameters, a precise determination of $\alpha_s$ based on a measurement of $R_{3/2}$ seems infeasible.

## Jet substructure

Jet shapes are dictated by multi-gluon emission from the primary outgoing parton with smaller contributions from initial state radiation and the underlying event. A measurement of the jet energy distribution as function of the distance to the jet axis allows to distinguish between quark and gluon jets (with the latter being wider) and provides constraints on event generator tunes [20].

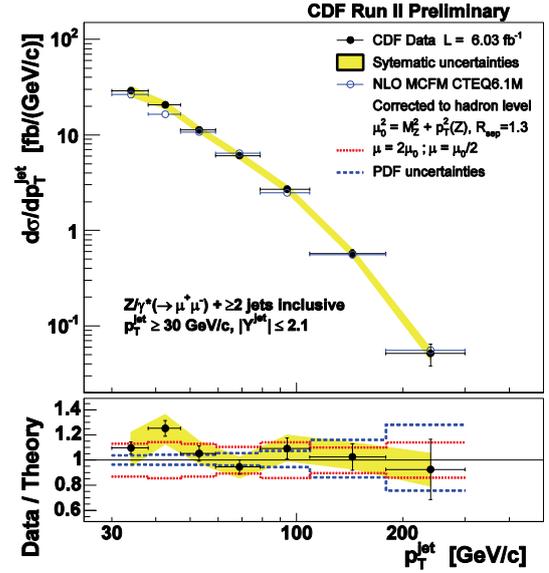

**Fig. 8:** *Measured inclusive jet differential cross section as a function of the jet $p_T$ in $Z+\geq 2$ jet events compared to NLO pQCD predictions [24].*

A further observable sensitive to the jet substructure is the jet mass. The jet algorithm used in Run II combines the jet components by adding their four-momenta (E-scheme), which consequently produces massive jets. Highly massive jets are expected, when e.g. the decay products of a high-$p_T$ top quark, $W$, $Z$ or Higgs boson are reconstructed within a single jet. Therefore, it is important to determine the rate of massive boosted jets from ordinary QCD jet production. A recent preliminary measurement by the CDF collaboration of the jet mass distribution in an inclusive jet sample based on 5.95 $fb^{-1}$ found a good agreement with the Pythia prediction [21].

## Associated production of jets with $W$ or $Z$ bosons

The production of $W$ or $Z$ bosons with associated jets is a critical background for top quark measurements and searches for the Higgs boson, supersymmetry and other new phenomena.

Both, the CDF and D0 collaborations measured the jet multiplicity and the jet $p_T$ distributions in $W/Z$+jets events using data sets with tion of $W/Z$+jets events, the Alpgen [27] and Sherpa [17] event generators, which combine LO matrix-element calculations (with up to 6 partons associated with the boson) with parton showers using dedicated matching algorithms are commonly used.

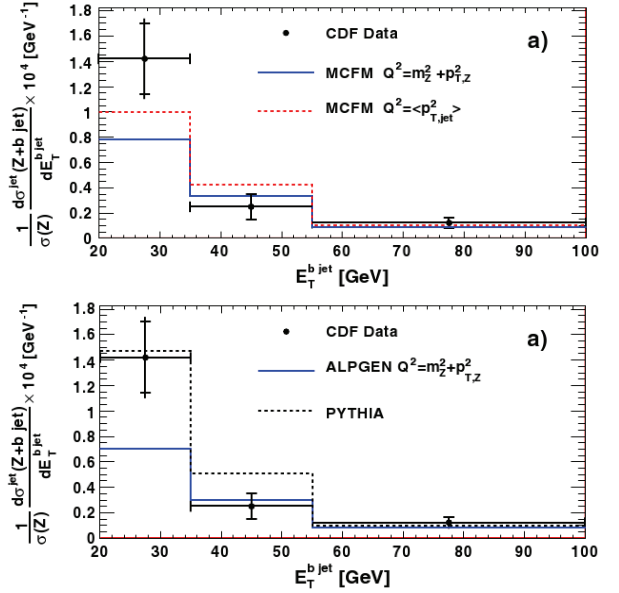

*Fig. 10:* *Measured differential cross section as function of the $E_T$ of the b-jet in Z+b jet events compared to the predictions of MCFM using two different scales (top) and Alpgen and Pythia (bottom) [30].*

The D0 collaboration compared the predictions of these generators as well as those from the traditional parton shower generators Herwig [25] and Pythia [16] with the measured jet $p_T$ distributions of the first, second, and third jet in Z+jet events (Fig. 9) [26]. The shape of the jet $p_T$ distributions are well described by Alpgen and Sherpa, but the scale uncertainties on the normalization of the predictions are large as the simulation is based on LO matrix-elements. Therefore, a validation of these event generators with data is crucial. Whereas the parton shower approach of Pythia and Herwig predict a too soft $p_T$ spectrum for non-leading jets, the leading jet is better modelled as the first or hardest, respectively, parton emission is modified using weights from matrix element calculations.

The D0 collaboration studied in addition angular observables, e.g. the distance in azimuthal angle and rapidity between the Z boson and the leading jet and found a reasonable description by the Sherpa generator [29].

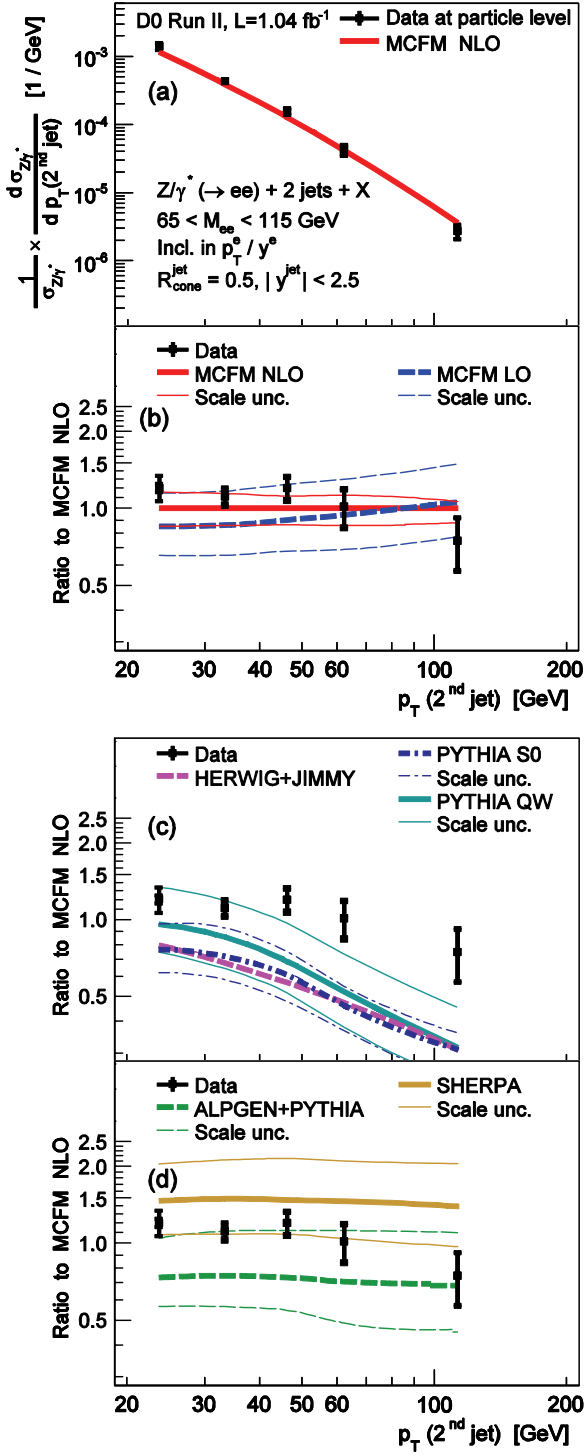

*Fig. 9:* *Measured differential cross section as function of the $p_T$ of the second leading jet in $Z+ \geq 2$ jet events (a) compared to the predictions of LO and NLO pQCD (b), Herwig and Pythia (c), and Sherpa and Alpgen (interfaced to Pythia) (d) [26].*

up to 6 $fb^{-1}$ of integrated luminosity [22-26]. In general, a good description of the data with the predictions of NLO pQCD was found (e.g. see Fig. 8). Only recently the NLO corrections for $W$+3 jet and $Z$+3 jet production were derived and processes with higher jet multiplicities can only be calculated at LO. For the simula-

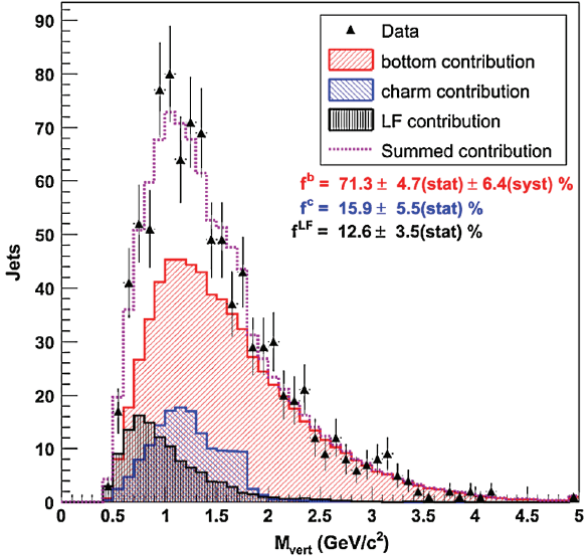

**Fig. 11:** *The CDF measurement of the vertex mass distribution for W+b jet candidate events compared to the maximum likelihood fit using templates for the bottom, charm, and light flavour (LF) contribution [33].*

## Associated production of heavy flavour with *W, Z* bosons or photons

Of particular interest is the production of heavy flavour jets with the electroweak bosons, not only since $Wb\bar{b}$ and $Zb\bar{b}$ are the dominant background sources for the search of a low mass Higgs boson at the Tevatron, but also since several processes are sensitive to the heavy flavour content of the proton structure.

Both the CDF and D0 collaborations published updated measurements of the production of $Z$ bosons with an associated $b$ jet using integrated luminosities of 2 $\text{fb}^{-1}$ and 4.2 $\text{fb}^{-1}$, respectively [30,31]. These measurements are sensitive to both $Z + b$ production, which probes the $b$ quark PDF, as well as $Z + b\bar{b}$ production, in which a gluon splits into a $b\bar{b}$ pair.

For jets with $E_T > 20$ GeV and $|\eta| < 1.5$, the CDF collaboration measured a normalized $Z + b$ production cross section of $\frac{\sigma(Z+b)}{\sigma(Z+\text{jet})} = (2.08 \pm 0.33(\text{stat.}) \pm 0.34(\text{syst.}))\%$ [30] in agreement with the order $\alpha_s^2$ prediction of 1.8-2.2% obtained with MCFM [32]. Fig. 10 shows the $E_T$ distribution of the $b$ jet compared to the predictions of MCFM, Pythia and Alpgen. The MCFM prediction, which is NLO for the $Zb$ contribution but only LO for the $Zb\bar{b}$ and $Zbg$ processes, shows a sizeable scale dependence and slightly underestimates the data at low $E_T$. In this kinematic region, Pythia, which predicts a rate about twice as large as Alpgen, is in better agreement with the data. This is primarily due to the different factorization and renormalization scales used by the two generators.

The D0 collaboration obtained a normalized $Z + b$ cross section of $\frac{\sigma(Z+b)}{\sigma(Z+\text{jet})} = (1.93 \pm 0.22(\text{stat.}) \pm 0.15(\text{syst.}))\%$ for jets with $E_T > 20$ GeV and within a wider region of $|\eta| < 2.5$ [31] in good agreement with the MCFM prediction of $(1.72 \pm 0.22)\%$.

Based on a fit to the vertex mass distribution (Fig. 11) the CDF collaboration measured the $W + b$ jet cross section to be $\sigma(W + b) \times Br(W \to l\nu) = (2.74 \pm 0.27(\text{stat.}) \pm 0.42(\text{syst.}))$ pb within the kinematic region defined by the charged lepton ($p_T^l > 20$ GeV, $|\eta^l| < 1.1$), neutrino ($p_T^\nu > 25$ GeV) and $b$ jet ($E_T^b > 20$ GeV, $|\eta^b| < 2$) [33]. The NLO QCD prediction of $(1.22 \pm 0.14)$ pb [34] is about $3\sigma$ too small and the prediction of Alpgen (interfaced to Pythia) of 0.78 pb is even lower.

$W + c$ production is dominated by the $s + g$ fusion process with a contribution of about 90% and thus provides constraints on the $s$ and $g$ PDFs at high $Q^2$.

For $c$ jets with $p_T^c > 20$ GeV and $|\eta^c| < 2$ the D0 collaboration measured a normalized production cross section of $\frac{\sigma(W+c)}{\sigma(W+\text{jet})} = (7.4 \pm 1.9(\text{stat.})^{+1.2}_{-1.4}(\text{syst.}))\%$ using a data set of 1 $\text{fb}^{-1}$ [32] compared to the prediction of Alpgen (interfaced to Pythia) of 4.4%.

A preliminary measurement by the CDF collaboration using 4.3 $\text{fb}^{-1}$ of data [36] found a $W + c$ cross section of $\sigma(W + c) \times Br(W \to l\nu) = (33.7 \pm 11.4(\text{stat.}) \pm 4.7(\text{syst.})$ pb for $c$ jets with $pTc > 12$ GeV and $|\eta^c| < 1.5$ compared to the NLO prediction of $(17.8 \pm 1.7)$ pb. While the predictions are lower than both measurements, they are in reasonable agreement given the substantial experimental uncertainties.

The production of photons with associated $b$ or $c$ jets is also sensitive to the heavy flavour and gluon content of the proton. While for $\gamma + b$ production, both the D0 and CDF collaborations observed a good agreement between their measurements and the NLO prediction over the full photon $p_T$ range [37,38], for $\gamma + c$ production the D0 collaboration found that the prediction underestimates the rate at large $p_T$ [37]. The description was slightly improved when PDF sets with inclusive charm contribution were used in the QCD calculation.

## Conclusions

Over the last years, our understanding of high $E_T$ and $W/Z$+jet production significantly advanced thanks to numerous precision measurements obtained at the Tevatron. These data benefit in particular from the precise jet energy scale calibrations.

The measurements of inclusive jet cross sections provide stringent constraints on the gluon PDF at high $x$ and allow a precise measurement of the running of $\alpha_s$. The cross sections for dijet and trijet production were found to be in agreement with NLO pQCD, albeit the data prefer the lower bound of the theory prediction. The dijet mass and angular distributions are sensitive to various extensions of the standard model and limits on e.g. quark compositeness and extra dimensions were set. Recently, the ATLAS and CMS experiments at LHC have started to improve these limits using the same experimental observables.

Measurements of the associated production of jets with the electroweak bosons are essential to test pQCD, to help generator modelling and tuning, and to constrain one of the dominant backgrounds to searches for the Higgs boson and supersymmetry, both at the Tevatron and LHC. For $W + b$ and $\gamma + c$ production some tension with theory was observed which still needs to be resolved.

Finally, both experiments at the Tevatron expect about 12 $fb^{-1}$ of data until the end of 2011 (and potentially more thereafter) which will allow to extend the measurements to higher jet $p_T$, dijet masses, and multiplicities.